\documentclass[preprint, amsmath,amssymb, aps,prl,showpacs]{revtex4-1}
\usepackage{epsfig}
\usepackage{dcolumn}% Align table columns on decimal point
\usepackage{bm}% bold math
\usepackage{tikz}
\usepackage{graphicx, graphics, color}
\usepackage{dcolumn}% Align table columns on decimal point
\usepackage{bm}% bold math
\usepackage{hyperref}% add hypertext capabilities
\usepackage[mathlines]{lineno}% Enable numbering of text and display math
\newcommand{\be}{\begin{equation}}
\newcommand{\ee}{\end{equation}}
\newcommand{\ben}{\begin{eqnarray}}
\newcommand{\een}{\end{eqnarray}}

\newcommand{\la}{{\lambda}}

\newcommand{\cK}{{\cal K}}

\newcommand{\cO}{{\cal O}}

\newcommand{\cL}{{\cal L}}
\newcommand{\cH}{{\cal H}}

\newcommand{\cM}{{\cal M}}

\newcommand{\cU}{{\cal U}}

\newcommand{\p}{\partial}
\newcommand{\na}{\nabla}

\newcommand{\hSi}{{\hat \Sigma}}

\newcommand{\hg}{\hat g}
\newcommand{\hR}{\hat R}

\newcommand{\tg}{\tilde g}

\newcommand{\tR}{{\tilde R}}

\newcommand{\tna}{\tilde \na}

\begin{document}

\title{Classification of static black holes in Einstein phantom/dilaton Maxwell/anti-Maxwell gravity systems}
%%%%%%%%%%%%%%%%%%%%%%%%%%%%%%%%%%%%%%%%%%%%%%%%%%%%%%%%%%%%%%%%%%%%%%%%%%%%%%%%%%%%%%%%%%%%%%%%%%%%%%%%%%%%%
\author{Marek Rogatko} 
\email{rogat@kft.umcs.lublin.pl}
%,marek.rogatko@poczta.umcs.lublin.pl }
%\author{Karol I. Wysoki\'nski}
%\email{karol@tytan.umcs.lublin.pl}
\affiliation{Institute of Physics, % \protect \\
Maria Curie-Sklodowska University, % \protect \\
20-031 Lublin, pl.~Marii Curie-Sklodowskiej 1, Poland}

\date{\today}% It is always \today, today,

%%%%%%%%%%%%%%%%%%%%%%%%%%%%%%%%%%%%%%%%%%%%%%%%%%%%%%%%%%%%%%%%%%%%%%%%%%%%%%%%%%%%%%%%%%%%%%%%%%
\begin{abstract}
The uniqueness theorem for static, spherically symmetric, asymptotically flat, higher dimensional phantom black holes, 
%%%%%
with non-degenerate event horizon
%%%%%%%%%
, being the solutions of Einstein
phantom/dilaton Maxwell/anti-Maxwell gravity systems is considered. Conformal positive energy theorem and conformal transformations authorize
 the crucial tools for exploiting the boundary conditions and conformal
flatness.
\end{abstract}

%\pacs{04.20.Jb, 04.40.-b}% PACS, the Physics and Astronomy
% 04.20.-q 	Classical general relativity
% 04.20.Fy 	Canonical formalism, Lagrangians, and variational principles
% 72.20.My 	Galvanomagnetic and other magnetotransport effects
\maketitle

%%%%%%%%%%%%%%%%%%%%%%%%%%%%%%%%%%%%%%%%%%%%%%%%%%%%%%%%%%%%%%%%%%%%%% 
\section{Introduction}
Recent astronomical and astrophysical observations provide sustenance for the fact that our universe consists of significant amount of non-baryonic
{\it dark matter} and the other mysterious ingredient of the Universe mass, causing its acceleration, {\it dark energy}. Due to the observations of cosmic microwave background
radiation conducted by Planck satellite and the experiments measuring the supernovae Type 1a distances, it is revealed that the {\it dark sector} constitutes, respectively
for {\it dark energy} and {\it dark matter}, almost $68$ and $27$ percent of its total mass  \cite{planck}. 
The expansion of the Universe can be mimicked by scalar fields with negative pressure, the so-called {\it phantom } fields. 
The model comprises the case of kinetic terms of the fields, with the 'wrong sign', in
comparison to the ordinary ones, which gives repulsive coupling to gravity. 
%%%%%%%%%
%{\color{red}
One should emphasize that the {\it phantom} fields violating null energy conditions, is one of the possibilities explaining the Universe expansion. In general the violation of the strong energy condition constitutes the enough factor. The future observations have to decide about the true nature of {\it dark energy}.
%%%%

As was revealed in  \cite{gib88}-\cite{cle05}, the presence of the additional scalar field in gravity systems, modified the known black hole solutions in highly non-trivial way.
On the other hand, in Ref. \cite{bro06} the problem of static, spherically symmetric  black holes in Einstein-Maxwell-dilaton gravity, with a phantom coupling was elaborated.
These new classes of black hole solutions with single or multiple event horizons, have also unusual causal structure.

Further, the generalization of the aforementioned studies was presented \cite{cle09} and the precise classification of black hole solutions in the theory in question was given.
Among all black hole spacetime with an infinite series of regular event horizons was found. In the studies the geometrically complete black hole solution was also revealed.

In \cite{azr11} the static multicenter solutions in phantom Einstein-Maxwell gravity were paid attention to and the regular black hole solutions without 
spatial symmetry for certain discrete values of
dilaton coupling were discovered. The three-dimensional gravitating sigma model being the result of dimensional reduction of phantom Einstein-Maxwell, phantom Kaluza-Klein,
and phantom Einstein-Maxwell-dilaton-axion theories were discussed.

Both analytical and numerical studies, revealed the importance of the influence of {\it dark energy}-phantom fields, in gravitational collapse \cite{cai06}-\cite{nak15}.
The various possible scenarios are possible, when one takes into account dark sector component coupling to the electrically charged scalar field.
For instance, the gravitational collapse subject to the presence of {\it dark energy} ensues the emergence of dynamical wormholes and naked singularities \cite{nak12, nak15}.

As far as the higher-dimensional spacetime is concerned, static spherically solutions of Einstein, Einstein-Maxwell-dilaton equations  with massless phantom field,
for $n \ge 4$ dimensional manifolds have been considered in \cite{noz20a}-\cite{noz20b}. It was found that they could be classified in three groups, i.e., the Fisher, the Ellis-Gibbons and 
the Ellis-Bronikov one. It happens that they constitute seeds for generating asymptotically A(dS) solutions \cite{noz20c}.

The possible justification of the existence of phantom fields one can seek in the string theory,  where they arise quite naturally in the studies of the so-called {\it negative tension branes}, e.g.,
the symmetry like $SU(N/M)$ can be realized like two stacks of branes, $N$-ordinary and $M$ of the negative tensions \cite{vaf01}. 
 Furthermore, in string theories, we also encounter the so-called ghost condensations, which in turn can precede to the phantom-like fields \cite{gas01}.
On the other hand, such kind of fields may in principle support traversability of wormhole solutions, in four-dimensional and their higher dimensional theories of gravity \cite{mor88}-\cite{tor13}. 
Their classification in four-dimensional wormhole and higher dimensional cases, have been studied recently in several works \cite{rub89}-\cite{rog18a}.

%%%%%%%%%%%%%%%%%%%%%%%%%%%%%%%%
Having in mind interesting features of the phantom black holes and phantom fields, as well as, the recent  astrophysical estimations of the abundance of {\it dark energy}
in our Universe, it will be not a miss to ask a question concerning the uniqueness of such kind of black hole solutions.

The motivation of our work is to explore the problem of black hole classification (uniqueness theorem) for phantom theories with $U(1)$-gauge field, with non-trivial
coupling among them. We shall elaborate the gravity theory provided by the following action:
\be
S = \int d^n x \sqrt{-g}~ \Big( R - 2 \eta_1 ~\na_\mu \phi \na^\mu \phi  + \eta_2 ~e^{\la \phi} F_{\mu \nu} F^{\mu \nu} \Big),
\label{act}
\ee
where $R$ stands for the Ricci scalar of $n$-dimensional manifold, $\na_\alpha$ denotes the Levi-Civita connection in the spacetime in question.
On the other hand, $F_{\mu \nu}$ represents $U(1)$-gauge field strength tensor, while $\phi$ is connected with dilaton one. $\la$ depicts coupling constant between gauge and scalar fields.
The action in question is string theory inspired one and has been widely studied from the point of view of the possible black hole/ wormhole solutions \cite{cle09, azr11}.
The two parameters $\eta_1$ and $\eta_2$ are equal to $\pm 1$, respectively. They enable one to study the case of Einstein dilaton ( $\eta_1 =1$), phantom ($\eta_1 =-1$), 
Maxwell ($\eta_2 = -1$), anti-Maxwell ($\eta_2 =1$)
systems.

%%%%%%%%%%%%%%%%%%%%%%%%%%%%%%%%%%%%%%%%%%%%%%%%%%%%%%%%%%%%%%%%%%%
\section{Classification - uniqueness theorem for phantom black holes}
In this section we shall provide the uniqueness theorem for static, spherically symmetric,
%%%%%%
asymptotically flat black objects with non-degenerate event horizons,
%%%%%%%
 in higher-dimensional gravity system described by the action (\ref{act}).
For the ordinary Einstein-Maxwell-dilaton gravity, being the low-energy limit of the heterotic string theory, the uniqueness for asymptotically
flat black hole objects comprises rather complicated mathematical challenge \cite{mas93}-\cite{yaz11}, in which
the key role in the proof, plays the conformal positive energy theorem \cite{sim99} and the adequate
the conformal transformations enabling. The conformal transformations enables one to examine the boundary conditions and the
conformal flatness of the spacetime under inspection.

To commence with let us suppose that the spacetime under inspection is 
static in the strict sense, having a timelike Killing vector field $\xi_\alpha = (\p/\p t)_\alpha$ defined at each point of the manifold in question. The definition of
staticity yields that the timelike Killing vector field is orthogonal to the $(n-1)$-dimensional hypersurface. It implies that 
the line element of the considered spacetime is provided by
\be
ds^2 = - V^2(x_i) dt^2 + g_{ij} dx^i dx^j,
\ee
where we set  $g_{ij}$ for the metric tensor of $(n-1)$-dimensional Riemanian manifold. Moreover, one also
imposes the staticity conditions for the fields appearing in the considered gravity theory. 
Thus, for the Maxwell field and phantom scalar they are written in the forms as follows:
\be
\cL_{\xi} F_{\mu \nu} = 0, \qquad \cL_\xi \phi = 0,
\ee
where  $\cL_\xi $ stands for the Lie derivative with respect to the Killing vector field $\xi$.

For the system in question the dimensionally reduced equations of motion are given by
\ben \label{rr}
{}^{(n-1)}R_{ij}&-&\frac{1}{V} {}^{(g)} \na_i {}^{(g)} \na_j V = 2\eta_1 {}^{(g)} \na_i  \phi {}^{(g)} \na_j \phi \\ \nonumber \label{np}
&+&2\eta_2 e^{2\la \phi} \Big[ \frac{{}^{(g)} \na_i \psi {}^{(g)} \na^i \psi}{V^2} + g_{ij} \frac{ {}^{(g)} \na_k \psi {}^{(g)} \na^k \psi}{(2-n) V^2} \Big], \\
{}^{(g)} \na_i {}^{(g)} \na^i \phi &+& \frac{{}^{(g)} \na_i V {}^{(g)} \na_i \phi}{V} \\ \nonumber
&+& \frac{\la~\eta_2}{\eta_1}~ e^{2 \la \phi} \frac{{}^{(g)} \na_k \psi {}^{(g)} \na^k \psi}{V^2} = 0,\\
&{}& {}^{(g)} \na_i \Bigg( \frac{e^{2 \la \phi} {}^{(g)}\na^i \psi}{V} \Bigg) = 0,\\
{}^{(g)} \na_i {}^{(g)} \na^i V &+& 2\eta_2~ \frac{e^{2 \la \phi} (n-3) {}^{(g)} \na_k \psi {}^{(g)} \na^k \psi}{(n-2) V} = 0, 
\een 
where by ${}^{(n-1)}R_{ij}$ and ${}^{(g)} \na_i $ we have denoted the Ricci scalar curvature and the 
%%%%%
%{\color {red} 
covariant derivative
%%%%%%
existing in $(n-1)$-dimensional manifold.
 $\psi$ describes the electrostatic potential.

%The strictly static spacetime assumption enables us to elaborate the situation when
%no event horizons are present in the considered manifold. 
%In order to get rid of singularities in  $(n-1)$-dimensional hypersurfaces ${}^{(n-1)}\Sigma $ of constant time, we assume 
 %that the $(n-1)$-dimensional submanifold is complete.

 Let us assume
further, that in asymptotically flat spacetime,
for a compact subset $\cK \subset {}^{(n-1)}\Sigma$, which is diffeomorphic to $R^{n-1}/B^{n-1}$, where $B^{n-1}$
is closed unit ball situated at the origin of $R^{n-1}$.
%%%%
 It implies that
%%%%
one has a standard coordinate system enabling the expansion as follows:
\ben \label{aa1}
g_{ij} &=& \bigg( 1 + \frac{2}{n-3} \frac{M}{r^{n-3}} \bigg) \delta_{ij} + \cO \Big(\frac{1}{r^{n-2}}\Big),\\
V &=& \bigg(1 - \frac{M}{r^{n-3}} \bigg) +  \cO \Big(\frac{1}{r^{n-2}} \Big),\\
\psi &=& \frac{Q}{r^{n-3}} + \cO \Big(\frac{1}{r^{n-2}} \Big),\\ \label{aa3}
\phi &=& \phi- \frac{q}{(n-3) r^{n-3}} + \cO \Big(\frac{1}{r^{n-2}} \Big),
\een
where $\phi,~M,~Q,~q$ are constant.
$M$ and $q$ represent the ADM masses and charges $Q$, defined up to a constant factor, while $r^2 = x_m x^m$.
The standard notion of asymptotically flat regions are provided by
relations (\ref{aa1})-(\ref{aa3}). 

%%%%%%%%%%%%%%%%%%%%%%%%%%%%%%%%%%%%%%%%%%%%%%%%%%%%%%%%%%%%%%%
The conformal positive energy theorem, derived in Ref. \cite{sim99}, will constitute the key role in the subsequent proof of the black hole uniqueness.
In order to apply it to the considerations we should satisfy its assumptions, i.e., one has to have two asymptotically flat 
Riemannian $(n-1)$-dimensional manifolds, $(\Sigma^{(\Phi)},~ {}^{(\Phi)}g_{ij})$ and $(\Sigma^{(\Psi)},~ {}^{(\Psi)}g_{ij})$,
which metric tensors are connected by the conformal transformation of the form
\be
{}^{(\Psi)}g_{ij} = \Omega^2~{}^{(\Phi)}g_{ij}, 
\ee
where $\Omega$ is a conformal factor. It turns out that the masses of the above manifolds fulfil the relation
${}^{(\Phi)}m + \beta~{}^{(\Psi)}m \geq 0$, under the additional requirement imposed on the Ricci scalar tensor
${}^{(\Phi)} R + \beta~\Omega^2~{}^{(\Psi)} R \geq 0$, 
where ${}^{(\Phi)} R $ and ${}^{(\Psi)} R$ are the Ricci scalars with respect to the adequate metric tensors, defined on the two manifolds.
$\beta$ is a positive constant. The inequalities in question are satisfied if
$(n-1)$-dimensional manifolds are flat \cite{sim99}.
The conformal positive energy theorem was widely use in proofs of the uniqueness of four and higher-dimensional black objects \cite{mar02}-\cite{yaz11},~
\cite{rog03}-\cite{rog13}, as well as
wormhole solutions \cite{rog18,rog18a}.

%%%%%%%%%%%%%%%%%%%%%%%%%%%%%%%%%%
To proceed to the uniqueness proof, let us define $(n-1)$-dimensional metric tensor which yields
\be
{}^{(n-1)}\tg_{ij} = V^{\frac{2}{n-3}} ~g_{ij}.
\label{crm}
\ee
The conformally rescaled metric tensor (\ref{crm}), implies that the Ricci curvature tensor has the form as follows:
\ben  \label{rij} 
&{}& {}^{(n-1)} \tR(\tg)_{ij} = \frac{1}{V^2} \Big( \frac{n-2}{n-3} \Big) {}^{(n-1)}\tna_i V{}^{(n-1)}\tna_j V \\ \nonumber
&+& 2\eta_1 {}^{(n-1)}\tna_i \phi {}^{(n-1)}\tna_j \phi + 2 \eta_2 e^{\la \phi} \frac{{}^{(n-1)}\tna_i \psi {}^{(n-1)}\tna_j \psi }{V^2}.
\een
In the next step, we define the quantities provided by the relations
\ben
\Phi_{\pm 1}&=& \frac{1}{2} \Bigg[ e^{C\phi} V \pm \frac{e^{- C\phi}}{V}
-\frac{D^2~e^{-C\phi} ~\psi^2}{V} \Bigg],\\
\Phi_0 &=&\frac{ D~e^{-C \phi}~\psi}{V},\\
\Psi_{\pm 1} &=& \frac{1}{2} \Bigg[
e^{- A\phi} \pm \frac{e^{A \phi}}{V} \Bigg],
\een
where $A= C/(n-3)$ and the constants $C,~D$ and $\la$ are bounded with the adequate values of $\eta_1$ and $\eta_2$ appearing in the action (\ref{act}).
For the brevity of notation we depict their exact values in Table I.
\begin{center}
\begin{table}
%\caption{Values of the constants for $\eta_i$}
\begin{tabular}{|c||c||c|}
\hline 
$\eta_i$ & $\eta_1 =1$ & $\eta_1 =-1$ \\
\hline \hline
$\eta_2 =1$ & $C=\sqrt{\frac{2 \eta_1}{n-2}}~ (n-3)$  & $ C=\sqrt{\frac{-2 \eta_1}{n-2}} ~(n-3)$ \\
&$D= i \sqrt{2 \eta_2 ~(n-3)}$ & $D= i \sqrt{2 \eta_2 ~(n-3)}$ \\
&$\la = - \sqrt{\frac{2 \eta_1}{n-2}} ~(n-3)$ & $\la = - \sqrt{\frac{-2 \eta_1}{n-2}} ~(n-3)$ \\
\hline
\hline
$\eta_2 =- 1$ &
$C=\sqrt{\frac{2 \eta_1}{n-2}}~ (n-3) $  & $ C=\sqrt{\frac{-2 \eta_1}{n-2}} ~(n-3)$ \\
&$D= \sqrt{2 \eta_2 ~(n-3)}$ & $D=  \sqrt{2 \eta_2 ~(n-3)}$ \\
&$\la = - \sqrt{\frac{2 \eta_1}{n-2}} ~(n-3)$ & $\la = - \sqrt{\frac{-2 \eta_1}{n-2}}~ (n-3)$ \\
\hline
\hline
\end{tabular}
\caption{Values of the constants for $\eta_i,~i =1,~2.$}
\end{table}
\end{center}

Then, the following symmetric tensors can be constructed on the aforementioned manifolds
\ben \nonumber \label{ric}
{}^{(\Phi)} \tR_{ij} &=& {}^{(n-1)}\tna_i \Phi_{-1} {}^{(n-1)}\tna_j \Phi_{-1} - {}^{(n-1)}\tna_i \Phi_0 {}^{(n-1)}\tna_j \Phi_0 \\ 
&-& {}^{(n-1)}\tna_i \Phi_1{}^{(n-1)}\tna_j \Phi_1, \\ \nonumber
{}^{(\Psi)} \tR_{ij} &=& {}^{(n-1)}\tna_i \Psi_{-1} {}^{(n-1)}\tna_j \Psi_{-1} - {}^{(n-1)}\tna_i \Psi_1 {}^{(n-1)}\tna_j \Psi_1.
\een

Setting metric in the given form $\eta_{AB} = diag(1,-1,-1)$, enables one to find that
$\Psi_{A} \Psi^{A} = \Phi_{A} \Phi^A = -1$, where $A= (0,~1,-1)$. Next, by virtue of the relation (\ref{ric}), one obtains
\ben \label{ps}
{}^{(n-1)}\tna_m {}^{(n-1)}\tna^m \Psi_B  &=& {}^{(\Psi)} \tR_{i}{}{}^{i} ~\Psi_B, \\ \label{ph}
{}^{(n-1)}\tna_m {}^{(n-1)}\tna^m \Phi_B  &=& {}^{(\Phi)} \tR_{i}{}{}^{i} ~\Phi_B.
\een
It can be also proved that 
the Ricci curvature tensor of the conformally rescaled metric ${}^{(n-1)}\tg_{ij} $ implies
\be
\tR_{ij} = \Big( {}^{(\Phi)} \tR_{ij}  + \frac{1}{n-3} {}^{(\Psi)} \tR_{ij} \Big).
\ee
%%%%%%%%%%%%%%%%%%%%%%%%%%%%%%%%%%%%%%%%%%%%%%%%%%%%%
Consequently, In order to meet
the requirements of the conformal positive energy theorem, one defines
the conformal transformations given by
\be
{}^{(\Phi)}g_{ij}^{\pm} = {}^{(\Phi)}\omega_{\pm}^{\frac{2}{n-3}}~ \tg_{ij},
\qquad
{}^{(\Psi)}g_{ij}^{\pm} = {}^{(\Psi)}\omega_{\pm}^{\frac{2}{n-3}}~ \tg_{ij},
\label{pff}
\ee
where the conformal factors yield
\be
{}^{(\Phi)}\omega_{\pm} = \frac{\Phi_{1} \pm 1 }{ 2}, \qquad
{}^{(\Psi)}\omega_{\pm} = \frac{\Psi_{1} \pm 1}{  2}.
\label{pf}
\ee
The careful scrutiny of the metric tensors defined by
the relation (\ref{pff}), envisages that we obtain four
$(n-1)$-dimensional manifolds, denoted respectively as  $(\Sigma^{+ (\Phi)},~ {}^{(\Phi)}g_{ij}^{+})$,
$(\Sigma^{- (\Phi)},~ {}^{(\Phi)}g_{ij}^{-})$, $(\Sigma^{+ (\Psi)},~ {}^{(\Psi)}g_{ij}^{+})$,
$(\Sigma^{- (\Psi)},~ {}^{(\Psi)}g_{ij}^{-})$. 

Pasting them together \cite{gib02,gib02a}, across the surface $V=0$, one achieves complete regular hypersurfaces
$\Sigma^{(\Phi)} = \Sigma^{+ (\Phi)} \cup \Sigma^{- (\Phi)}$ and $\Sigma^{(\Psi)} = \Sigma^{+ (\Psi)} \cup \Sigma^{- (\Psi)}$, and moreover
${}^{(\Phi)}g_{ij}^{\pm}$ and ${}^{(\Phi)}g_{ij}^{\pm}$ metrics are complete. 

The asymptotic conditions imposed on $g_{ij}$, electric potential $\psi$, and scalar field, show their explicit asymptotical behavior.

The resulting manifolds $\Sigma^{(\Phi)} $ and $\Sigma^{(\Psi)} $ are geodesically complete.
If $( \Sigma^{(m)},~{}^{(m)}g_{ij},~\Phi_A,~\Psi_A)$, where $m= \Phi,~\Psi$, are asymptotically flat solution of the equations (\ref{ps}) and (\ref{ph}), with non-degenerate black hole event horizon,
in the next step one ought to check if the gravitational mass on them is equal to zero.

 %%%%%%%%%%%%%%%
 In the next step it remains to show that the static slice is conformally flat. In order to
this goal, we implement the conformal positive energy theorem \cite{sim99}.
Let us define the other conformal transformation described by
\be
\hg^{\pm}_{ij} = \Big[ \Big( {}^{(\Phi)}\omega_{\pm} \Big)^{\frac{2}{n-3}}
 \Big( {}^{(\Psi)}\omega_{\pm} \bigg)^{2} \Big]^{\frac{1}{n-2}} \tg_{ij}.
\ee
On the other hand, 
the Ricci curvature tensor on the defined space, may be written as
\ben \nonumber \label{ricci}
&{}&\Big[ \Big( {}^{(\Phi)}\omega_{\pm} \bigg)^{\frac{2}{n-3}}
 \bigg( {}^{(\Psi)}\omega_{\pm} \bigg)^{2} \Big]^{\frac{1}{n-2}}~\hR^\pm
\\
&=& \Big( {}^{(\Phi)}\omega_{\pm}^{\frac{2}{n-3}} ~{}^{(\Phi)}R^\pm + \frac{1}{n-3}~
{}^{(\Psi)}\omega_{\pm}^{\frac{2}{n-3}}~ {}^{(\Psi)}R^\pm \bigg) 
\\ \nonumber
&+& \frac{1}{n-3}~
\Big( \tna _{i} \ln {}^{(\Phi)}\omega_{\pm} - \tna _{i} \ln {}^{(\Psi)}\omega_{\pm} \Big)^2. 
\een
Consequently, by the direct calculations
it can be verified that
the first term in the brackets in the relation (\ref{ricci}), may be rewritten as follows:
\ben \nonumber \label{ricci1}
&{}&{}^{(\Phi)}\omega_{\pm}^{\frac{2}{n-3}} ~{}^{(\Phi)}R^\pm + \frac{1}{n-3}~
{}^{(\Psi)}\omega_{\pm}^{\frac{2}{n-3}}~ {}^{(\Psi)}R^\pm \\ \nonumber
&=&
\frac{\mid { \Phi_{0} \tna_{i} \Phi_{-1} - \Phi_{-1} \tna_{i} \Phi_{0} } \mid^2}{(\Phi_1 \pm1)^2} \\ 
&+& \frac{1}{n-3}~
\frac{\mid { \Psi_{1} \tna_{i} \Psi_{-1}
- \Psi_{-1} \tna_{i} \Psi_{1} } \mid^2}{(\Psi_1 \pm 1)^2}.
\een
The relations (\ref{ricci}) and (\ref{ricci1}) reveal the conclusion that $\hR^\pm$ is greater or equal to zero. 
Then, by virtue of the application of the conformal positive energy theorem, one has that
$(\Sigma^{(\Phi)},~ {}^{(\Phi)}g_{ij})$, $(\Sigma^{(\Psi)},~ {}^{(\Psi)}g_{ij})$ and
$(\hSi,~ \hg_{ij})$ are flat and these facts entail that
the conformal factors
${}^{(\Phi)}\omega = {}^{(\Psi)}\omega$ and $\Phi_{1} = \Psi_{1}$, as well as,
$\Phi_{0} = const~ \Phi_{-1}$ and $\Psi_{0} = const~ \Psi_{-1}$. 

Finally one can  conclude that the manifold $({}^{(n-1)} \Sigma,~ g_{ij})$ is conformally flat and the metric tensor $\hg_{ij}$ may be depicted in a 
conformally flat form. Namely, we define a function provided by the following:
\be
\hg_{ij} = {\cal U}^{\frac{4}{n-3}}~ {}^{(\Phi)}g_{ij},
\label{gg}
\ee
where we set ${\cal U} = ({}^{(\Phi)}\omega_{\pm} V)^{-1/2}$.\\
The considered equations of motion of the studied gravity system reduce now
to the Laplace equation, on the three-dimensional Euclidean manifold. Namely we have
\be
\na_{i}\na^{i}{\cal U} = 0,
\ee
where $\na$ is the connection on a flat manifold. It follows from the fact that 
the Ricci scalar for the metric $\hg_{ij}$ is equal to zero.
%%%%%%%%%%%%%%%%%%%%
Moreover one has that the following metric for the flat base space is valid
\be
{}^{(\Phi)}g_{ij} = {\tilde \rho}^{2} d{\cU}^2 + {\tilde h}_{AB}dx^{A}dx^{B},
\ee
where ${\tilde \rho}^2 = \na_b {\cU} \na^b {\cU}$.

%%%%%%%%%%
In the next step we shall demonstrate that the conformally transformed event horizon constitutes a geometric sphere. In order to proceed, let us consider
how the event horizon is embedded into the base space $(\hat \Sigma,~\hg_{ij})$. Namely
%%%%%%%%%%
, we can define a local coordinate in the neighborhood $\cM \in \hSi$, for the flat base space
\be
\hg_{ij}dx^{i}dx^{j} = \delta_{ij}dx^{i}dx^{j} = \rho^{2} d{\cU}^2 + { h}_{AB}dx^{A}dx^{B}.
\ee
The manifold in question is totally geodesic, which means that any of its sub-manifold geodesic is a geodesic in the considered manifold.
The event horizon is located at some $\cU = const$ and 
the other important fact is that the embedding of $\hSi$ into Euclidean $(n-1)$-manifold is totally umbilical \cite{kob69}, which results that each connected 
component of $\hSi$ constitutes a geometric sphere 
of a certain radius.  

The studied embedding is also rigid \cite{kob69}, which yields that
one is always able to locate one connected component of the event horizon $\cH$, of a certain radius $\rho$, at $r=r_0$ surface on $\hSi$. Thus,
the above mathematical construction leads us to a boundary value problem for the Laplace equation on the base space $\Theta = E^{n-1}/B^{n-1}$, with a Dirichlet boundary conditions.

The system in question
is characterized by a parameter which fixes the radius of the inner boundary  and authorizes a black hole of a specific radius $\rho_{\mid _{\cH}}$ in the gravity theory described by the
action (\ref{act}). 
We have also the following limit condition  for $\cU$, i.e., it tends to ${\cU} \sim 1 + \cO(r^{n-3})$, when $r \rightarrow \infty$. 

Let us assume further, that we have two solutions being subject to the same boundary value problem, i.e., 
%%%%
$\cU_1$ and $\cU_2$.
%%%%%%%
By means of Green
identity and integrating over the volume $\Theta$, one has that
\ben \nonumber
\Big( \int_{r \rightarrow \infty} - \int_{\cH} \Big) (\cU_1- \cU_2)~\frac{\p}{\p r} (\cU_1-\cU_2) dS \\
= \int_\Theta \mid \na (\cU_1 - \cU_2) \mid^2 d \Theta.
\een
%%%%%
The surface integrals on the left-hand side of the above relation vanish due to the imposed boundary conditions. This fact provides that the volume  integral have to be identically equal to zero.
%%%%
We conclude that the 
aforementioned two solutions of the Laplace equation with the Dirichlet boundary conditions are identical. It accounts for the main conclusion of our considerations.

\noindent
{\bf Theorem:}\\
Let us assume that $\cU_1$ and $\cU_2$ constitute the two solutions of the Laplace equation on the base space $\Theta = E^{n-1}/B^{n-1}$, as defined above.
They authorize solutions of 
Einstein phantom/dilaton Maxwell/anti-Maxwell (depending on the special choice of $\eta_i$, where $i =1,~2$)
of equations of motion, describing static, spherically symmetric, asymptotically flat adequate black holes with
%%%%
non-degenerate event horizons.
%%%%%
The solutions of the equations of motion of the theory under inspection are subject to the same boundary and regularity conditions. Then, $\cU_1 = \cU_2$ in 
all of the region of the base space 
%%%%%%%
%{\color{red} 
$\Theta$,
%%%%%%%
provided that
$\cU_1(p) = \cU_2(p)$ for at least one point belonging to the aforementioned region.

%%%%%%%%%%%%%%%%%%%%%%%%%%%%%%%%%%%%%%%%%%%%%%%%%%%%%%%%%%%%%%%%%%%%%%%
\section{Conclusions}
The recent astrophysical and astronomical observations reveals that {\it dark energy} is a possible ingredient of our Universe, which implicate that
phantom fields and phantom black objects should be subject of interest and careful scrutiny. 
In our paper we have constructed the uniqueness theorem for 
Einstein phantom/dilaton Maxwell/anti-Maxwell, depending on the special choice on $\eta_i$, where $i =1,~2$, black hole solutions.

We have paid attention to $n$-dimensional spherically symmetric, 
%%%%%%
asymptotically flat
%%%%%%
, static black hole with non-degenerate event horizon, being the solution of the aforementioned theory. 
The key role in the proof was played by the conformal positive energy theorem
and conformal transformations helping us to examine the boundary conditions and the conformal flatness of the examined spacetime.

%%%%%%%%%%%%%%%%%%%%%%%%%%%%%%%%%%%%%%%%%%%%%%%%%%%%%%
%\begin{acknowledgments}
% {MR was partially supported by the grant DEC-2014/15/B/ST2/00089 of the National Science Center.}
% and KIW by the grant \mbox{DEC-2014/13/B/ST3/04451}.
%\end{acknowledgments}

%%%%%%%%%%%%%%%%%%%%%%%%%%%%%%%%%%%
%\end{document}
%%%%%%%%%%%%%%%%%%%%%%%%%%%%%%%%%%%%%%%%%%%%%%%%%%%%%%%%%%%%%%%%%%%%%%%%%%%%%%%%%
%%%%%%%%%%%%%%%%%%%%%%%%%%%%%%%%%%%%%%%%%%%%%%%%%%%%%%%%%%%%%%%%%%%

%%%%%%%%%%%%%%%%%%%%%%%%%%%%%%%%%%%%%%%%%%%%%%%%%%%%%%%%%%%%%%%%%%%%%%%%%%%%%%%%%%%%%%%%%%%%%%%%%%%%%%%%%%%%%%%%%%%%


\begin{thebibliography}{99}

%
\def\cmp#1#2#3#4{\emph{#4}, \emph{ Commun. Math. Phys.} {\bf #1} (#3) #2}
\def\lmp#1#2#3#4{\emph{#4}, \emph{ Lett. Math. Phys.} {\bf #1} (#3) #2}
\def\hpa#1#2#3#4{\emph{#4}, \emph{ Hell. Phys. Acta} {\bf #1} (#3) #2}
\def\grg#1#2#3#4{\emph{#4}, \emph{ Gen. Rel. Grav.} {\bf #1} (#3) #2}

%\def\pr#1#2#3#4{\emph{#4}, \emph{ Phys. Rev.} {\bf #1} #2 (#3)}

\def\pr#1#2#3#4{\emph{#4}, \emph{ Phys. Rev.} {\bf #1} (#3) #2}
\def\prl#1#2#3#4{\emph{#4}, \emph{ Phys. Rev. Lett.} {\bf #1} (#3) #2}
\def\prd#1#2#3#4{\emph{#4}, \emph{ Phys. Rev. D} {\bf #1} (#3) #2}
\def\pl#1#2#3#4{\emph{#4}, \emph{ Phys. Lett.} {\bf #1} (#3) #2}
\def\pla#1#2#3#4{\emph{#4}, \emph{ Phys. Lett. A} {\bf #1} (#3) #2}
\def\plb#1#2#3#4{\emph{#4}, \emph{ Phys. Lett. B} {\bf #1} (#3) #2}
\def\prep#1#2#3#4{\emph{#4}, \emph{ Phys. Reports} {\bf #1} (#3) #2}
\def\phys#1#2#3#4{\emph{#4}, \emph{ Physica} {\bf #1} (#3) #2}
\def\jcp#1#2#3#4{\emph{#4}, \emph{ J. Comput. Phys.} {\bf #1} (#3) #2}
\def\jmp#1#2#3#4{\emph{#4}, \emph{ J. Math. Phys.} {\bf #1} (#3) #2}
\def\jpm#1#2#3#4{\emph{#4}, \emph{ J. Phys. A: Math. Gen.} {\bf #1} (#3) #2}
\def\cpr#1#2#3#4{\emph{#4}, \emph{ Computer Phys. Rept.} {\bf #1} (#3) #2}
\def\cqg#1#2#3#4{\emph{#4}, \emph{ Class. Quant. Grav.} {\bf #1} (#3) #2}
\def\cma#1#2#3#4{\emph{#4}, \emph{ Computers Math. Applic.} {\bf #1} (#3) #2}
\def\mc#1#2#3#4{\emph{#4}, \emph{ Math. Compt.} {\bf #1} (#3) #2}
\def\apj#1#2#3#4{\emph{#4}, \emph{ Astrophys. J.} {\bf #1} (#3) #2}
\def\apjs#1#2#3#4{\emph{#4}, \emph{ Astrophys. J. Suppl.} {\bf #1} (#3) #2}
\def\acta#1#2#3#4{\emph{#4}, \emph{ Acta Astronomica} {\bf #1} (#3) #2}
%%%%%%%%%%%%%%%%%%%%%%%%%%%%%%%%%%%%%%%%%%%%%%%%%%%%%%%%%%%%%%%%%%%%%%%%%%
\def\apl#1#2#3#4{\emph{#4}, \emph{ Ann. Physik. (Leipzig)} {\bf #1} (#3) #2}
\def\amjp#1#2#3#4{\emph{#4}, \emph{Am. J. Phys.} {\bf #1} (#3) #2}
\def\anp#1#2#3#4{\emph{#4}, \emph{ Ann. Phys.} {\bf #1} (#3) #2}
\def\sa#1#2#3#4{\emph{#4}, \emph{ Sov. Astro.} {\bf #1} (#3) #2}
\def\sia#1#2#3#4{\emph{#4}, \emph{ SIAM J. Sci. Statist. Comput.} {\bf #1} (#3) #2}
\def\aa#1#2#3#4{\emph{#4}, \emph{ Astron. Astrophys.} {\bf #1} (#3) #2}
\def\mnras#1#2#3#4{\emph{#4}, \emph{ Mon. Not. R. Astr. Soc.} {\bf #1} (#3) #2}
\def\npb#1#2#3#4{\emph{#4}, \emph{ Nucl. Phys. B} {\bf #1} (#3) #2}
\def\prsla#1#2#3#4{\emph{#4}, \emph{ Proc. R. Soc. London, Ser. A} {\bf #1} (#3) #2}
\def\jhep#1#2#3#4{\emph{#4}, \emph{ JHEP} {\bf #1} (#2) #3}
\def\jcap#1#2#3#4{\emph{#4}, \emph{ J. Cosmol. Astropart. Phys.} {\bf #1} (#2) #3}
\def\nuc#1#2#3#4{\emph{#4}, \emph{ Nuovo Cimento B } {\bf #1} (#3) #2}
\def\ijmp#1#2#3#4{\emph{#4}, \emph{ Int. J. Mod. Phys. D} {\bf #1} (#3) #2}
\def\atmp#1#2#3#4{\emph{#4}, \emph{ Adv. Theor. Math. Phys.} {\bf #1} (#3) #2}
\def\ptps#1#2#3#4{\emph{#4}, \emph{ Prog. Theor. Phys. Suppl.} {\bf #1} (#3) #2}
\def\lmp#1#2#3#4{\emph{#4}, \emph{ Lett. Math. Phys.} {\bf #1} (#3) #2}
\def\cpam#1#2#3#4{\emph{#4}, \emph{ Comm. Pure Appl. Math.}  {\bf #1} (#3) #2}
\def\adv#1#2#3#4{\emph{#4}, \emph{ Adv. Phys.}  {\bf #1} (#3) #2}
\def\zh#1#2#3#4{\emph{#4}, \emph{ Zh. Eksp. Teor. Fiz.}  {\bf #1} (#3) #2}

\def\jams#1#2#3#4{\emph{#4}, \emph{ J. Austral. Math. Soc. B} {\bf #1} (#3) #2}
\def\appa#1#2#3#4{\emph{#4}, \emph{ Acta Phys. Polonica A} {\bf #1}, (#3) #2}
\def\nat#1#2#3#4{\emph{#4}, \emph{Nature} {\bf #1}, (#3) #2}
\def\science#1#2#3#4{\emph{#4}, \emph{Science} {\bf #1}, (#3) #2}
\def\arcmp#1#2#3#4{\emph{#4}, \emph{Annual Rev. of Cond. Matter Physics} {\bf #1}, (#3) #2}
\def\jcap#1#2#3#4{\emph{#4}, \emph{JCAP} {\bf #1}, (#3) #2}
\def\conphy#1#2#3#4{\emph{#4}, \emph{Contemporary Physics} {\bf #1}, (#3) #2}
\def\ptps#1#2#3#4{\emph{#4}, \emph{ Prog. Theor. Phys. Suppl.} {\bf #1} (#3) #2}
\def\ptp#1#2#3#4{\emph{#4}, \emph{ Prog. Theor. Phys.} {\bf #1} (#3) #2}
\def\apjsup#1#2#3#4{\emph{#4}, \emph{ Astrophys. J. Suppl. Ser.} {\bf #1} (#3) #2}
\def\mplb#1#2#3#4{\emph{#4}, \emph{ Mod. Phys. Lett. B} {\bf #1} (#3) #2}
\def\ijmpd#1#2#3#4{\emph{#4}, \emph{ Int. J. Mod. Phys. D} {\bf #1} (#3) #2}

\def\gravcos#1#2#3#4{\emph{#4}, \emph{ Grav. Cosmol.} {\bf #1} (#2) #3}
\def\ajp#1#2#3#4{\emph{#4}, \emph{ Am. J. Phys.} {\bf #1} (#3) #2}
\def\appb#1#2#3#4{\emph{#4}, \emph{ Acta Phys. Polon. B} {\bf #1} (#3) #2}
%
\def\hepph#1#2{{ hep-ph }{#1} (#2)}
\def\hepth#1#2{{ hep-th }{#1} (#2)}
\def\astroph#1#2{{ astro-ph }{#1} (#2)}
\def\grqc#1#2{{ gr-qc }{#1} (#2)}
\def\ibid#1#2#3#4{\emph{#4}, {\it ibid.} {\bf #1} (#3) #2}

%
%%%%%%%%%%%%%%%%%%%%%%%%%%%%%%%%%%%%%%%%%%%%%%%%%%%%%%%%%%%%%%%%%%%%%%%%%%%%%%%%%
%%%%%%%%%%%%%%%%%%%%%%%%%%%%%%%%%%%%%%%%%%%%%%%%%%%%%%%%%%%%%%%%%%%%%%%%%%%%%%%%%
%%%%%%%%%%%%%%%%%%%%%%%%%%%%%%%%%%%%%%%%%%%%%%%%%%%%%%%%%%%%%%%%%%%%%
\bibitem{planck}
Planck collaboration, P.A.R.  Ade et al., \aa{571}{A16}{2014}{Planck 2013 results. XVI. Cosmological parameters}.

\bibitem{gib88}
G.W. Gibbons and K. Maeda, \npb{298}{741}{1988}{Black holes and membranes in higher-dimensional theories with dilaton fields}.
\bibitem{gar91}
D. Garfinkle, G.T. Horowitz, and A. Strominger, \prd{43}{3140}{1991}{Charged black holes in string theory}.
\bibitem{cle03}
G. Clement, C. Leygnac, and D. Gal'tsov, \prd{67}{024012}{2003}{Linear dilaton black holes}.
\bibitem{cle04}
G. Clement and C. Leygnac, \prd{70}{084018}{2004}{Non-asymptotically flat, non-AdS dilaton black holes}.
\bibitem{cle05}
G. Clement, C. Leygnac, and D. Gal'tsov, \prd{71}{084014}{2005}{Black branes on the linear dilaton background}.


%%%%%%%%%%%%%%%PH BH %%%%%%%%%%%%%%%%%%%%%%%
%\bibitem{gib96}
%G.W. Gibbons and D.A. Rasheed, \npb{476}{515}{1996}{Dyson pairs and zero mass black holes}.

%\bibitem{cle99}
%G. Clement and A. Fabbri, \cqg{16}{323}{1999}{The gravitating  $\sigma$ model in 2 +1 dimensions: black hole solutions}.
\bibitem{bro06}
K. A. Bronnikov and J. C. Fabris, \prl{96}{251101}{2006}{Regular phantom black holes}.
%\bibitem{bro98}
%K.A. Bronikov, C.P. Constantinidis, R.L. Evangelista, and J.C. Fabris, \ijmpd{8}{481}{1999}{Electrically charged cold black holes in scalar-tensor theories}.
%\bibitem{bro98a}
%K.A. Bronikov, G. Clement, C.P. Constantinidis, and J.C. Fabris, \pla{243}{121}{1998}{Structure and stability of cold scalar-tensor black holes}.
\bibitem{cle09}
G. Clement, J.C. Fabris, and M.E. Rodrigues, \prd{79}{064021}{2009}{Phantom black holes in Einstein-Maxwell-dilaton theory}.
\bibitem{azr11}
M. Azreg-Ainou, G. Clement, J.C. Fabris, and M.E. Rodrigues, \prd{83}{124001}{2011}{Phantom black holes and sigma models}.



%%%%%%%%%%%%%%%%%%%%%%DARK ENERGY COLLAPSE%%%%%%%%%%%%%%%%%%%%%%%%%%%%%%%%%%%%%%%%%
\bibitem{cai06}
R.-G. Cai and A. Wang, \prd{73}{063005}{2006}{Black hole formation from collapsing dust fluid in a background of dark energy}.
\bibitem{cha10}
S. Chakraborty and T. Bandyopadhyay, \gravcos{16}{151}{2010}{Collapse dynamics of a star of dark matter and dark energy}.
\bibitem{wan09}
Q. Wang and Z. Fan, \prd{79}{123012}{2009}{Dynamical evolution of quintessence dark energy in collapsing dark matter halos}.
\bibitem{del13}
M. Le Delliou and T. Barreiro, \jcap{02}{2013}{037}{Interacting dark energy collapse with matter components separation}.
\bibitem{nak12}
A. Nakonieczna, M. Rogatko,  and R. Moderski, \prd{86}{044043}{2012}{Dynamical collapse of charged scalar field in phantom gravity}.
 \bibitem{nak15}
A. Nakonieczna, M. Rogatko,  and L. Nakonieczny, \jhep{11}{2015}{012}{Dark sector impact on gravitational collapse of an electrically charged scalar field}.


\bibitem{noz20a}
C. Martinez and M. Nozawa, \hepth{2010.05183}{2010}~{\it Static spacetimes haunted by a phantom scalar field: classification and global structure in the massless case}.
\bibitem{noz20b}
M. Nozawa, \hepth{2010.07560}{2010}~{\it Static spacetimes haunted by a phantom scalar field II: dilatonic charged solutions}.
\bibitem{noz20c}
M. Nozawa, \hepth{2010.07561}{2010}~{\it Static spacetimes haunted by a phantom scalar field III:  asymptotic (A)dS solutions}.


\bibitem{vaf01}
C. Vafa, \hepth{01011218}{2001},~{\it Brane/anti-brane systems and $U(N/M)$ supergroup}.
\bibitem{gas01}
M. Gasperini, F. Piazza, and G. Veneziano, \prd{65}{023508}{2001}{Quintessence as a runaway dilaton}.





















%%%%%%%%%%%%%%%%%%%%%%%WORMHOLES%%%%%%%%%%%%%%%%%%%%%%%%%%%%%%%%%%%%%%%
\bibitem{mor88}
M.S. Morris, K.S. Thorne, and U. Yurtsever, \prl{61}{1446}{1988}{Wormholes, time machines and weak energy condition}.
\bibitem{ell73}
H.G. Ellis, \jmp{14}{104}{1973}{Ether flow through a drainhole: A particle model in general relativity}.
\bibitem{bro73}
K. Bronikov, \appb{4}{251}{1973}{Scalar-Tensor Theory and Scalar Charge}.
%\bibitem{ell79}
%H.G.Ellis, \grg{10}{105}{1979}{The evolving, flowless drainhole: A nongravitating-particle model in general relativity theory}.
\bibitem{mor88amj}
M.S. Morris and K.S. Thorne, \ajp{56}{395}{1988}{Wormholes in spacetime and their use for interstellar travel: A tool for teaching general relativity}.
%\bibitem{kan11}
%P. Kanti, B. Kleinhaus, and J. Kunz, \prl{107}{271101}{2011}{Wormholes in Dilatonic Einstein-Gauss-Bonnet Theory}.
\bibitem{teo98}
E. Teo, \prd{58}{024014}{1998}{Rotating tranversable wormholes}.
\bibitem{cle84}
G. Clement, \grg{16}{131}{1984}{A class of wormhole solutions to higher-dimensional general relativity}.
\bibitem{gib16}
G.W. Gibbons and M.S. Volkov, \plb{760}{324}{2016}{Ring wormholes via duality rotations}.
\bibitem{gib17}
G.W. Gibbons and M.S. Volkov, \jcap{05}{039}{2017}{Weyl metrics and wormholes}.
\bibitem{gou18}
P. Goulart, \cqg{35}{025012}{2018}{Phantom wormholes in Einstein-Maxwell-dilaton theory}.

\bibitem{bro97}
K.A. Bronikov and J.C. Fabris, \cqg{14}{831}{1997}{Weyl spacetimes and wormholes in D-dimensional Einstein and dilaton gravity}.
\bibitem{deb03}
A. DeBenedictis and A. Das, \npb{653}{279}{2003}{Higher dimensional wormhole geometries with compact dimensions}.
\bibitem{tor13}
T. Tori and H. Shinkai, \prd{88}{064027}{2013}{Wormholes in higher dimensional space-time: Exact solutions and their linear stability analysis}.


\bibitem{rub89}
P.J. Ruback, \cqg{6}{L21}{1989}{A uniqueness theorem for wormholes in quantum gravity}.
\bibitem{yaz17}
S. Yazadjiev, \prd{96}{044045}{2017}{Uniqueness theorem for static wormholes in Einstein phantom scalar field theory}.
\bibitem{laz17}
B. Lazov, P. Nedkova, and S. Yazadjiev, \plb{778}{408}{2018}{Uniqueness theorem for static phantom wormholes in Einstein-Maxwell-dilaton theory}.
\bibitem{rog18}
M. Rogatko, \prd{97}{024001}{2018}{Uniqueness of higher-dimensional phantom field wormholes}.
\bibitem{rog18a}
M. Rogatko, \prd{97}{064023}{2018}{Uniqueness of higher-dimensional Einstein-Maxwell-phantom dilaton field wormholes}.






%\bibitem{rogwoj18}
%M. Rogatko and A. Wojnar, \jcap{05}{2018}{023}{Dark matter influence on black objects thermodynamics}.






%%%%%%%%%%%%%%%%%%%%%%%%
\bibitem{mas93}
A.K.M. Massod-ul-Alam, \cqg{10}{2649}{1993}{Uniqueness of static charged dilaton black holes}.
\bibitem{rog99}
M. Rogatko, \prd{59}{104010}{1999}{Uniqueness theorem for static dilaton $U(1)^2$ black holes}.
\bibitem{rog02}
M. Rogatko, \cqg{19}{875}{2002}{Uniqueness theorem for static dilaton $U(1)^N$ black holes}.
\bibitem{mar02}
M. Mars and W. Simon, \atmp{6}{279}{2002}{On uniqueness of static Einstein-Maxwell black holes}.
\bibitem{rog16}
M. Rogatko, \prd{93}{064003}{2016}{Uniqueness of photon sphere for Einstein-Maxwell-dilaton black holes with arbitrary coupling constant}.
\bibitem{gib02}
G.W. Gibbons, D. Ida, and T. Shiromizu, \prl{89}{041101}{2002}{Uniqueness and nonuniquess of static black holes in higher dimensions}.
\bibitem{gib02a}
G.W. Gibbons, D. Ida, and T. Shiromizu, \prd{66}{044010}{2002}{Uniqueness of (dilatonic) charged black holes and black p-branes in higher dimensions}.
\bibitem{yaz11}
S. Yazadjiev, \jhep{06}{2011}{083}{Uniqueness and non-uniqueness of the stationary black holes in 5D Einstein-Maxwell and Einstein-Maxwell dilaton gravity}.







%%%%%%%%%%%%%%%%%%%%%%%%%%%%%%%%%%%
\bibitem{sim99}
W. Simon, \lmp{50}{275}{1999}{Conformal Positive Mass Theorems}.

%%%%%%%%%%%%UNIQ BH %%%%%%%%%%%%%%
%\bibitem{gib02}
%G.W.Gibbons, D.Ida, and T.Shiromizu, \prl{89}{041101}{2002}{Uniqueness and nonuniquess of static black holes in higher dimensions}.
%\bibitem{gib02a}
%G.W.Gibbons, D.Ida, and T.Shiromizu, \prd{66}{044010}{2002}{Uniqueness of (dilatonic) charged black holes and black p-branes in higher dimensions}.
\bibitem{rog03}
M. Rogatko, \prd{67}{084025}{2003}{Uniqueness theorem of static degenerate and nondegenerate charged black holes in higher dimensions}.
\bibitem{rog04}
M. Rogatko, \prd{70}{044023}{2004}{Uniqueness theorem for generalized Maxwell electric and magnetic black holes in higher dimensions}.
 \bibitem{rog06}
 M. Rogatko, \prd{73}{124027}{2006}{Classification of static charged black holes in higher dimensions}.
\bibitem{rog13}
M. Rogatko, \prd{88}{024051}{2013}{Uniqueness of charged static asymptotically flat black holes in dynamical Chern-Simons gravity}.




%\bibitem{posen}
%R.Schoen and S.T.Yau, \cmp{65}{45}{1979}{On the proof of the positive mass conjecture in general relativity},\\
%E.Witten, \cmp{80}{381}{1981}{A new proof of the positive energy theorem}.
\bibitem{kob69}
S. Kobayashi and K. Nomizu, {\it Foundations of Differential Geometry}, (Intersciene, New York, 1969), vol.II.



%\bibitem{yaz15}
%S.Yazadjiev, \prd{91}{123013}{2015}{Uniqueness of static spactimes with a photon sphere in Einstein-scalar field theory}.
%\bibitem{bun87}
%G.L.Bunting and A.K.M.Masood-ul-Alam, \grg{19}{147}{1987}{Nonexistence of multiple black holes in asymptotically Euclidean static vacuum space-time}.




%%%%%
\end{thebibliography}
\end{document}